\documentclass[aps,prl,twocolumn,showpacs,superscriptaddress,groupedaddress]{revtex4}

\usepackage{bbm}

\usepackage{graphicx}
\usepackage{dcolumn}
\usepackage{amsmath}
\usepackage{bm}
\usepackage{color}
\usepackage{mathrsfs}
\usepackage{epstopdf}
\usepackage{amsmath}
\usepackage{amssymb}
\usepackage{subfigure}
\usepackage{enumitem}
\usepackage{multirow}
\usepackage{exscale}
\usepackage{relsize}
\usepackage{dsfont}

\newcommand{\be}{\begin{equation}}
\newcommand{\ee}{\end{equation}}
\newcommand{\bey}{\begin{eqnarray}}
\newcommand{\eey}{\end{eqnarray}}
\newcommand{\bw}{\begin{widetext}}
\newcommand{\ew}{\end{widetext}}

\newcommand{\ba}{\begin{array}}
\newcommand{\ea}{\end{array}}
\newcommand{\bi}{\begin{itemize}}
\newcommand{\ei}{\end{itemize}}
\newcommand{\bem}{\begin{enumerate}}
\newcommand{\eem}{\end{enumerate}}

\begin{document}

\title{Work statistics across a quantum phase transition}

\author{Zhaoyu Fei}
\affiliation{School of Physics, Peking University, Beijing 100871, China}

\author{Nahuel Freitas}
\affiliation{Complex Systems and Statistical Mechanics, Physics and Materials Science,
University of Luxembourg, L-1511 Luxembourg, Luxembourg}

\author{Vasco Cavina}
\affiliation{Complex Systems and Statistical Mechanics, Physics and Materials Science,
University of Luxembourg, L-1511 Luxembourg, Luxembourg}

\author{H. T. Quan} \email[Email: ]{htquan@pku.edu.cn}
\affiliation{School of Physics, Peking University, Beijing 100871, China}
\affiliation{Collaborative Innovation Center of Quantum Matter, Beijing 100871, China}
\affiliation{Frontiers Science Center for Nano-optoelectronics, Peking University, Beijing, 100871, China}

\author{Massimiliano Esposito} \email[Email: ]{massimiliano.esposito@uni.lu}
\affiliation{Complex Systems and Statistical Mechanics, Physics and Materials Science,
University of Luxembourg, L-1511 Luxembourg, Luxembourg}

 \date{\today}

\begin{abstract}
We investigate the statistics of the work performed during a quench across
a quantum phase transition using the adiabatic perturbation theory.
It is shown that all the cumulants of work exhibit universal scaling behavior analogous to the Kibble-Zurek scaling for the average density of defects.
Two kinds of transformations are considered: quenches between two gapped phases
in which a critical point is traversed, and quenches that end near the critical point.
In contrast to the scaling behavior of the density of defects,
the scaling behavior of the work cumulants
are shown to be qualitatively different for these two kinds of quenches. However,
in both cases the corresponding exponents are fully determined by the
dimension of the system and the critical exponents of the
transition, as in the traditional Kibble-Zurek mechanism (KZM). Thus, our study
deepens our understanding about the nonequilibrium dynamics of
a quantum phase transition by revealing the imprint of the KZM on the work statistics.
\end{abstract}

\maketitle

\textit{Introduction}.---
In cosmology and condensed matter physics the creation of excitations during
continuous phase transitions (thermal or quantum) is usually described by the
Kibble-Zurek mechanism (KZM)~\cite{kibble1976,kibble1980,zurek1985,zurek1996}. The KZM relates the \emph{average}
density $\langle n_\text{ex} \rangle$ of excitations or defects created during a transformation or quench
across a critical point to the rate or speed at which the critical region is
traversed.
This is particularly relevant for adiabatic quantum computation
and simulation schemes, where non-adiabatic effects impose a tradeoff between
the speed and the fidelity that can be achieved~\cite{kim2010, biamonte2011, albash2018}.
Importantly, the KZM predicts a universal power law dependence of $\langle n_\text{ex} \rangle$ on this rate, with an exponent that is fully determined
by the dimension of the system
and the critical exponents of the transition. Of course, the
actual number of excitations created during a particular realization of the quench
is a stochastic quantity that will fluctuate from one realization to the next, and thus
must be characterized by a probability distribution. The traditional heuristic argument behind
the KZM, as well as more rigorous derivations based on the adiabatic perturbation theory~\cite{no2011,le2010,un2005},
only gives information about the first moment of this distribution, i.e., the average
density of excitations $\langle n_\text{ex}\rangle$.
However, it was recently shown by del Campo that in the exactly
solvable one dimensional (1D) transverse Ising chain the universal scaling predicted by the KZM also applies
to all the cumulants of $n_\text{ex}$~\cite{delcampo2018}.

Motivated by this finding, we extend previous results in two important aspects.
In the first place, we turn our attention away from the density of created excitations
and focus instead on the amount of work applied during the quench. Thus, we investigate
what are the signatures of the KZM on the characteristic function of work (CFW),
which plays an important role in the newly developed field of stochastic thermodynamics~\cite{st2010,eq2011,st2012}.
In analogy to the partition function, which  contains essential information about an equilibrium state, the CFW contains essential information about an arbitrary nonequilibrium process.
This interesting quantity
has received much attention since it allows to understand the emergence of
irreversibility during a thermodynamic transformation (via the fluctuation relations
~\cite{st2008, dorner2012}), and is related to other interesting quantities employed to study
the non-equilibrium dynamics of complex many-body systems like the Loschmidt echo
~\cite{st2008, dy2013, de2006, cr2011}.
Secondly, we provide a general scaling argument, underpinned by the well-known
results in the adiabatic perturbation theory~\cite{no2011,le2010,un2005}, showing that all the cumulants
of the work distribution exhibit a scaling behavior similar to the KZM scaling
for systems that can be described in terms of independent quasiparticles.
Our predictions are valid in principle for systems in
arbitrary dimensions, and are explicitly shown to hold in the exactly solvable
1D quantum transverse Ising model.


\textit{KZM and the adiabatic perturbation theory}.---
We first briefly review the basic concepts and heuristic arguments
behind the KZM scaling in a quantum phase transition, and also how to recover (and generalize) the same
results using the adiabatic perturbation theory.
We consider a second-order quantum phase transition between two gapped phases
characterized by the correlation length critical exponent $\nu$ and the dynamic critical exponent
$z$~\cite{halperin2019theory}. Thus, close to the quantum critical point, the
energy gap $\Delta$ between the ground state and the first relevant excited state,
the relaxation time $\tau$ and the correlation length $\xi$ scale
as~\cite{dy2010}
 \begin{gather}
  \label{e1}
  \Delta\sim|\lambda|^{z\nu}\qquad
  \tau\sim|\lambda|^{-z\nu}\qquad
  \xi\sim|\lambda|^{-\nu},
 \end{gather}
where $\lambda$ is a dimensionless parameter which measures the distance from the
critical point. We also consider a protocol in which the Hamiltonian of the
system is modified in such a way that the parameter $\lambda(t)$ can be
approximated as a linear quench $\lambda(t)=vt$ near the critical point, where $v>0$
is the quench rate.
The system is initially prepared in the ground state at $t_0\to -\infty$
and the protocol stops at $t_1\to\infty$. According to the KZM,
the evolution of the system can be divided into three parts: (1) $t_0<t<-t^*$,
(2) $-t^*<t<t^*$ and (3) $t^*<t<t_1$, where the time $t^*$ is determined
by the following argument~\cite{zurek1985,zurek1996,dy2010,dy2005}. During parts (1) and (3), the relaxation time
of the system is sufficiently small for its evolution to be considered adiabatic ($\tau < v^{-1}$),
since it can always catch up with the change of $\lambda(t)$ (adiabatic region). In contrast,
during part (2) the relaxation time is large compared to $v^{-1}$ and as a consequence
the state of the system is frozen out (impulse region). The freeze-out time $t^*$ can be estimated by the relation $t^* \simeq \tau(\lambda(t^*))$ and thus we obtain
$t^*\sim v^{-z\nu/(z\nu+1)}$. Then, the initial state for the adiabatic dynamics of
part (3) is approximately the final state of the evolution of part (1), and is
therefore characterized by a correlation length $\xi^* = \xi(\lambda(t^*)) \simeq v^{-\nu/(z\nu+1)}$.
This correlation length corresponds to the characteristic length of the system, e.g., the size of the magnetic domains.
Thus, the average density of defects or domain walls can be estimated as
\be \langle n_{ex} \rangle \sim \xi^{*-d}\sim v^{\frac{d\nu}{z\nu+1}}, \ee
where $d$ is the dimension of the system.


The above results can be reproduced by using the adiabatic perturbation theory~\cite{no2011,le2010,un2005}.
For this, we consider a system defined on a $d$-dimensional lattice and described by a Hamiltonian
$\hat H(\lambda(t))=\hat H_0+\lambda(t)\hat V$, where $\hat H_0$ is the
Hamiltonian at the quantum critical point and $\lambda(t)$, now called
the work parameter~\cite{eq2011}, is controlled by an external agent according to the above protocol.
Here, $\hat V$ is the driving Hamiltonian.
We assume that the system can be described by independent quasiparticles
(denoted by mode $k$), and that at the critical point the energy gap vanishes due to
the fact that the dispersion relation of low-energy (long wavelength, small $k$) modes exhibits
the scaling behavior $\omega_{k}=c|k|^z$ ($\hbar=1$)~\cite{halperin2019theory}, where $c$ is a non-zero constant.
We also assume that at most one low-energy quasiparticle can get excited after the quench,
which is called the few-excitation approximation in this article.
Then, within the adiabatic perturbation theory, the excitation
probability of the $k$th-mode quasiparticle $p_{k}$ is dominated by (assuming
that there is no additional Berry phase)~\cite{le2010,dy2010,un2005}
\be
\label{e2}
p_k\approx\left|\int_{\lambda_0}^{\lambda_1}\mathrm d\lambda \langle 1_k(\lambda)|\partial_{\lambda}|0_k(\lambda)\rangle e^{iv^{-1}\int_{\lambda_0}^{\lambda}\mathrm d\lambda'\omega_k(\lambda')}\right|^2,
\ee
where  $\partial_{\lambda}=\partial/\partial\lambda$, $\lambda_0=\lambda(t_0)$, $\lambda_1=\lambda(t_1)$ and $|n_k(\lambda)\rangle$ denotes the instantaneous energy eigenstate of mode $k$ of $\hat H(\lambda)$  with the occupation number $n_k$.
Then, the average density of excitations $\langle n_{ex} \rangle$ reads
\be
\langle n_{ex} \rangle =\lim_{N\to\infty}\frac{1}{N}\sum_{k}p_k=\int\frac{\mathrm{d}^dk}{(2\pi)^d}p_k,
\ee
where $N$ denotes the number of the lattice points.
In order to remove the quantity $v^{-1}$ in the exponential function in the
integral of $p_k$ (Eq.~(\ref{e2})), we introduce two rescaled quantities,
$\theta$ and $\phi$, defined by~\cite{le2010,dy2010,un2005}
\begin{equation}
\lambda=\theta\: v^{1/(z\nu+1)} \qquad k=\phi\: v^{\nu/(z\nu+1)}
\end{equation}
Also following Ref.~\cite{le2010,dy2010}, we introduce the general scaling argument
\begin{gather}
 \begin{split}
\omega_k(\lambda) &=|\lambda|^{z\nu} \: F(k/|\lambda|^{\nu}),\\
\langle 1^\lambda_k|\partial_{\lambda}|0^\lambda_k\rangle &=
\lambda^{-1}\: G(k/|\lambda|^{\nu}),
 \end{split}
\end{gather}
where $F$ and $G$ are two model-dependent scaling functions satisfying
$F(x)\propto x^z$ and $G(x)\propto x^{-1/\nu}$ for $|x|\gg1$. This is motivated
by dimensional considerations and the requirement that the spectrum of the
high energy modes should be insensitive to $\lambda$.
Thus, $\langle n_{ex}\rangle$ reads~\cite{le2010,un2005}
\be
\label{e4}
\langle n_{ex}\rangle \approx \:v^{\frac{d\nu}{z\nu+1}}\int\frac{\mathrm{d}^d\phi}{(2\pi)^d}K(\phi),
\ee
where
\be
\label{e7}
K(\phi)\!=\!\left|\int_{\theta_0}^{\theta_1}\!\frac{\mathrm d\theta}{\theta}G\left(\!\frac{\phi}{|\theta|^{\nu}}\!\right)\exp\left[i\!\int_{\theta_0}^{\theta}\mathrm \!\!d\theta'|\theta'|^{z\nu}F\left(\!\frac{\phi}{|\theta'|^{\nu}}\!\right)\right]\right|^2\!\!,
\ee
with $\lambda_0=\theta_0 v^{1/(1+z\nu)}$ and $\lambda_1=\theta_1 v^{1/(1+z\nu)}$.
For $d\nu/(z\nu+1)<2$, the integral Eq.~(\ref{e7}) converges in the limit $v\to 0$ and therefore
$\langle n_{ex} \rangle \sim v^{d\nu/(z\nu+1)}$, in accordance to the KZM prediction. But for $d\nu/(z\nu+1)>2$, the integral Eq.~(\ref{e7}) does not converge, which means that it is not dominated by the low energy modes~\cite{no2011,le2010,un2005}.
The high-energy (ultra-violet) contribution to the integral can be approximated by the regular analytic adiabatic perturbation theory~\cite{no2011,le2010,un2005}, which results in the quadratic scaling $\langle n_{ex}\rangle\sim v^2$ (see supplemental material). For $d\nu/(z\nu+1)=2$, an additional logarithmic correction is expected, i.e.,
$\langle n_{ex}\rangle\sim v^2\ln v$~\cite{no2011,le2010,sc2009}. This concludes our review of the KZM and the adiabatic perturbation theory. In the following, they are applied in analyzing the scaling
behaviour of the work distribution  during a linear quench.

\textit{Scaling behavior in the characteristic function of work}.---
We define the work applied during the quench on the basis of the usual
two-time measurement scheme, i.e., as the difference between the results of
the projective measurements~\cite{aq2000,ja2000,fl2007,esp2009, camp2011} of the system's energy before and after the linear quench.
It is a stochastic quantity with a distribution function $P(w)$, and the logarithm of its
characteristic function (the Fourier transform of $P(w)$), called the cumulant CFW,
reads
\begin{align}
\ln \chi(u)=&\ln\mathrm{Tr}[\hat{U}^\dag(t_1,\!t_0)e^{iu\hat H(\lambda_1)}\hat
 U(t_1,\!t_0)e^{-iu\hat H(\lambda_0)}\hat \rho_0] \nonumber \\
=&\sum_{n=1}^{\infty}\frac{(iu)^n}{n!}\kappa_n,
\label{e3}
\end{align}
where $\hat U(t_1,t_0)$ is the time evolution operator, $\hat \rho_0$ is the initial state and $\kappa_n$ is the $n$th cumulant of work~\cite{foot3}. We assume that $\lambda(t)$ evolves according to the above protocol, the system is initially prepared in the ground state of  $\hat H(\lambda_0)$ and the cumulant CFW satisfies a large deviation principle~\cite{th2009}, i.e., $\lim_{N\to\infty} N^{-1}\ln\chi(u)$ exists.

For the adiabatic driving ($v\to 0$), the system is in the ground state all the
time. Hence, we have $\kappa_1=\sum_k
[\varepsilon^0_k(\lambda_1)-\varepsilon^0_k(\lambda_0)]\equiv N\mu$, where
$\varepsilon^0_k(\lambda)$ denotes the zero-point energy of the $k$th mode of $\hat H(\lambda)$ and $\kappa_n=0$ for $n\geq2$ due to the definite measurement
results. Thus, the cumulant CFW for the adiabatic process should be $\ln\chi_a(u)=Niu\mu$
according to Eq.~(\ref{e3}). For nonadiabatic driving ($v$ is small but
nonzero), since the nonadiabatic corrections to the cumulant CFW come from the impulse
region of the KZM, we expect these corrections to exhibit the following scaling relation,
i.e., $\kappa_1=N(\mu+v^{\delta_1}f_1)$ and $\kappa_n=Nv^{\delta_n}f_n$, $n\geq
2$, where $f_n$ are model-dependent scaling functions and $\delta_n$ are the
corresponding exponents characterizing the scaling behavior of each cumulant.

Every exponent $\delta_n$ can be determined as follows. According to Eq.~(\ref{e3})
and in the few-excitation approximation, the scaling behavior of
$\kappa_n$ should be the scaling behavior of
$N^{-1}\sum_{k}[\omega_k(\lambda_1)]^np_k$ because the excitations of
quasiparticles in different modes are independent, i.e.,
$\kappa_1\approx N[\mu+N^{-1}\sum_{k}\omega_k(\lambda_1)p_k]$,
$\kappa_n\approx\sum_{k}[\omega_k(\lambda_1)]^np_k$ (see supplementary material). Now, by utilizing the expressions of $\kappa_1$ and $\kappa_n$ and following the same procedure as in the last section, we obtain
\begin{gather}
 \begin{split}
 \label{e10}
v^{\delta_n}f_n\: \approx \:\:&v^{\frac{d\nu}{z\nu+1}}\int\frac{\mathrm{d}^d\phi}{(2\pi)^d}[\omega_{k=\phi v^{\nu/(z\nu+1)}}(\lambda_1)]^nK(\phi).
 \end{split}
\end{gather}
If we fix $\lambda_0$ and $\lambda_1$ when varying $v$ and $\lambda_1$ is away from the critical point, $\omega_k(\lambda_1)$ is a non-zero constant when $k\to 0$. Also because only low-energy modes can be excited after the quench, we obtain
\begin{gather}
 \begin{split}
v^{\delta_n}f_n\: \approx
\:\:&[\omega_{k=0}(\lambda_1)]^nv^{\frac{d\nu}{z\nu+1}}\int\frac{\mathrm{d}^d\phi}{(2\pi)^d}K(\phi).
 \end{split}
\end{gather}
Following the same analysis as that after Eq.~(\ref{e4}), the exponents in the cumulant CFW read
\begin{equation}
\label{e13}
v^{\delta_n}=\left\{
\begin{aligned}
&v^{d\nu/(z\nu+1)}\ \ \ &d\nu/(z\nu+1)<2\\
&v^2\ln v  &d\nu/(z\nu+1)=2 \\
&v^2   &d\nu/(z\nu+1)>2.
\end{aligned}
\right.
\end{equation}
Finally, according to Eq.~(\ref{e3}), since in this case $\delta_n$ is independent of $n$, $\ln \chi(u)$ reads
\be
\label{e15}
\ln \chi(u)=\ln \chi_a(u)+Nv^{\delta_n}f(iu),
\ee
where $f(iu)\equiv\sum_{n=1}^{\infty}(iu)^nf_n/n!$. We would like to emphasize that the scaling behavior exists not only in the CFW, but also in the work distribution.
According to the
Gartner-Ellis theorem, the distribution of the work per lattice site $P_N(w)\equiv
P(Nw)$ also takes on the large deviation form, $\lim_{N\to\infty}N^{-1}P_N(w)=-I(w)$.
Here, the rate function $I(w)$ is obtained by the Legendre-Fenchel transform~\cite{th2009} via
\be
I(w)=\sup_{u\in\mathbb{R}}\{u(w-\mu)-v^{\delta_n}f(u)\}.
\ee

\begin{figure}
  \vspace{-.5cm}
  \includegraphics[scale=.465]{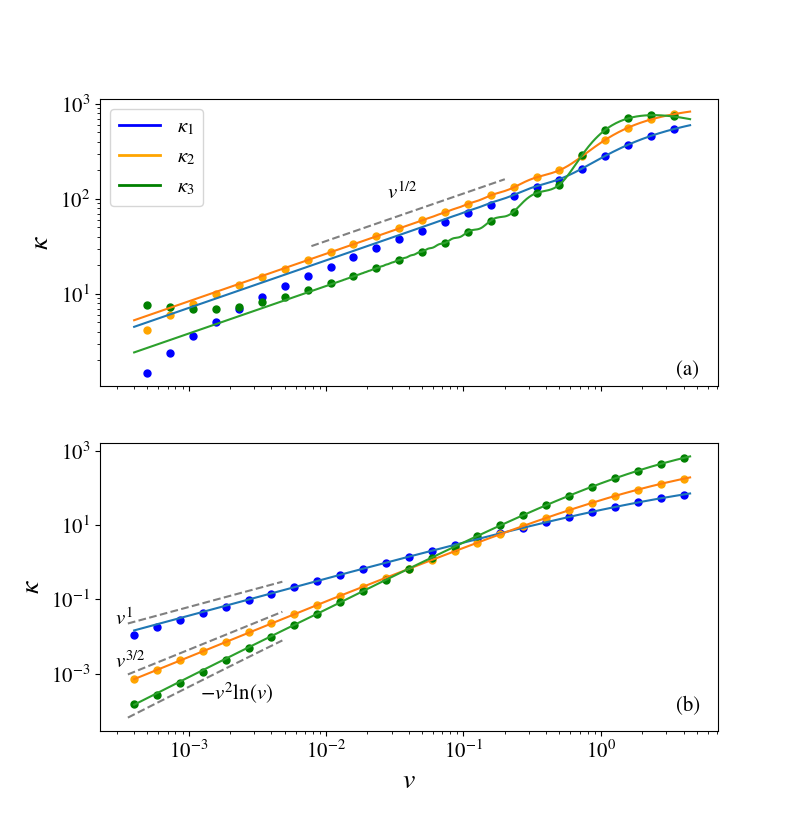}
  \vspace{-.5cm}
  \caption{The first three cumulants of the work distribution as a function of the
  quenching rate for a 1D transverse Ising chain. Solid lines correspond to the
  exact analytic solution in the macroscopic limit~\cite{foot4} and dots to an exact numerical simulation of a chain of $N=1000$ spins. (a) Quench between $\lambda_0=-4$ and $\lambda_1=1$, for which all the exponents are 1/2. The deviations at low $v$ are due to the finite size effects
  ~\cite{delcampo2018}. (b) Quench between $\lambda_0=-4$ and $\lambda_1=0$, for which our theory predicts $\kappa_1 \propto v$,
  $\kappa_2\propto v^{3/2}$ and $\kappa_3 \propto v^2 \ln v$.
  Finite size effects are less relevant in this case.}
  \label{fig:1d_ising_sim}
\end{figure}

Now, let us consider a second case: $\lambda_1$ is near the critical point. Since in this case $\omega_k(\lambda_1)=c|k|^z$ when $k\to 0$, according to Eq.~(\ref{e10}), we have
\begin{gather}
 \begin{split}
 \label{e12}
v^{\delta_n}f_n \: \approx
\:\:
&c^nv^{\frac{(d+nz)\nu}{z\nu+1}}\int\frac{\mathrm{d}^d\phi}{(2\pi)^d}|\phi|^{nz}K(\phi).
 \end{split}
\end{gather}
Similar to the discussion about $\langle n_{ex}\rangle$, we obtain
\begin{equation}
\label{e16}
v^{\delta_n}=\left\{
\begin{aligned}
&v^{(d+nz)\nu/(z\nu+1)}\ \ \ &(d+nz)\nu/(z\nu+1)<2\\
&v^2\ln v  &(d+nz)\nu/(z\nu+1)=2 \\
&v^2   &(d+nz)\nu/(z\nu+1)>2.
\end{aligned}
\right.
\end{equation}
We note that the quantity in Eq.~(\ref{e12})
for $n=1$ is called the excess energy in Refs.~\cite{no2011,le2010}.
If $\delta_1<2$, to a good approximation, we can cut off the sum in Eq.~(\ref{e3}) to the first order ($n=1$) for sufficiently small $v$ and obtain
\be
\ln \chi(u)\approx\ln \chi_a(u)+Nv^{\delta_1}iuf_1.
\ee
Accordingly, $P_N(w)$ is a Dirac delta distribution located at $\mu+v^{\delta_1}f_1$.

In summary, our analysis shows that the scaling of the work cumulants
is qualitatively different depending on whether $\lambda_1=0$.
If $\lambda_1\neq0$ all the
cumulants (for whatever $n$) have the same scaling exponent, while if $\lambda_1=0$ they do not.
This is illustrated in Figure~\ref{fig:1d_ising_sim} by the exact
numerical simulation of the 1D transverse Ising model~\cite{dyn2005}, which is
also studied analytically in the following. It is important to note that
this difference between the two kinds of quenches is not observed for the
density of excitations $n_\text{ex}$, which displays the same scaling behavior
irrelevant to the ending point of the protocol.

\textit{Example}.---We calculate the CFW of the 1D transverse Ising model to
demonstrate our results since it is solvable and the KZM is valid in
this model~\cite{dy2005,dyn2005}. The Hamiltonian of a chain of $N$ spins in a
transverse magnetic field reads
\be
\hat H(\lambda)=-J\sum_{l=1}^{N}[\hat \sigma_l^z\hat \sigma_{l+1}^z+(\lambda-1)\hat \sigma_l^x]
\ee
with the Born-von K\'{a}rm\'{a}n boundary condition. Here, $\hat \sigma^{x,y,z}_l$ denote the Pauli matrices on site $l$, and $J$ denotes the energy scale. The critical points are
at $\lambda=0,2$. Moreover, $d=z=\nu=1$~\cite{dy2010,dy2005}.
For the critical point $\lambda=0$, we choose $\lambda(t)=vt$,
$\lambda_0 < 0$ and $0<\lambda_1<2$. According to Ref.~\cite{gr2019}, when $N\to\infty$, the cumulant CFW  reads
\be
\label{e8}
\ln \chi(u)=\frac{N}{\pi}\int_{0}^{\pi}\mathrm{d}k\ln\frac{g_k(u)}{g_k(0)},
\ee
where
 \begin{gather}
  \begin{split}
  \label{e9}
g_k(u)=&\{1+\mathrm{cos}(u\omega_k^1)\mathrm{cos}[(u-i\beta)\omega_k^0]\\
&+Q_k\mathrm{sin}(u\omega_k^1)\mathrm{sin}[(u-i\beta)\omega_k^0]\}^{\frac{1}{2}}.
  \end{split}
 \end{gather}
Here, $\beta=(k_BT)^{-1}$ is the inverse temperature of the canonical initial
state, $\omega_k^0=\omega_k(\lambda_0),\omega_k^1=\omega_k(\lambda_1)$, where
$\omega_k(\lambda)=2J\sqrt{(\lambda-1+\mathrm{cos}k)^2+\sin^2 k}$ is the energy
of the $k$th mode. Also, $Q_k=1-2p_k$, where $p_k\approx e^{-2\pi J k^2/v}$ is
the excitation probability in the corresponding Landau-Zener model for mode
$k$~\cite{dyn2005}. From Eq.~(\ref{e8}), we obtain the first and the second
cumulants of work
  \begin{align}
  \label{e14}
\kappa_1\!=&\frac{N}{2\pi}\!\int_0^\pi \! \mathrm dk(\omega_k^0-Q_k\omega_k^1)\tanh{\left(\frac{\beta\omega_k^0}{2}\right)},\nonumber \\
\kappa_2\!=&\frac{N}{4\pi}\!\int_0^\pi \! \mathrm dk[(\omega_k^0-Q_k\omega_k^1)^2+(1-Q_k^2)(\omega_k^1)^2\cosh(\beta\omega_k^0)] \nonumber \\
&\times\mathrm{sech}^2\left(\frac{\beta\omega_k^0}{2}\right).
  \end{align}
Quantum phase transitions occur at the absolute zero. Hence, we consider the case in which the initial state is chosen to be the ground state of $\hat H(\lambda_0)$. From Eq.~(\ref{e8}), we have
  \begin{align}
\!\!\!&\ln \chi(u)=\!N\!\!\int_{0}^{\pi} \! \frac{\mathrm dk}{2\pi}\{iu(\omega_k^0-\omega_k^1)+\ln[1+p_k(e^{2iu\omega_k^1}-\!1)]\} \nonumber \\
\!\!\!&\!=\!\ln \chi_a(u)\!+\!N\!\sum_{n=1}^{\infty}\frac{(-1)^{n+1}}{n}\!\int_{0}^{\pi}\!\frac{\mathrm dk}{2\pi}p_k^n(e^{2iu\omega_k^1}-1)^n,
\label{e19}
\end{align}
where $\ln \chi_a(u)=Niu\mu=Niu(2\pi)^{-1}\int_{0}^{\pi}\mathrm
dk(\omega_k^0-\omega_k^1)$ is the cumulant CFW for the adiabatic process. The sum in the last equation is convergent
under the condition $|p_k(e^{2iu\omega_k^1}-1)|=2p_k|\sin(u\omega_k^1)|<1$~\cite{foot1}.
Also, for $T=0$, from Eq.~(\ref{e14}), we have
 \begin{gather}
  \begin{split}
&\kappa_1=N\left(\mu+\frac{1}{\pi}\int_0^\pi \mathrm dk \omega_k^1p_k\right),\\
&\kappa_2=\frac{2N}{\pi}\int_0^\pi\mathrm dk(\omega_k^1)^2p_k(1-p_k).
  \end{split}
  \label{eq:cumulants_zero_temp}
 \end{gather}
Due to the exponential decay of $p_k$, only low-energy modes can get excited.
Thus, we extend the upper limit of the integral to $\infty$ and approximate
$p_ke^{2iu\omega_k^1}\approx p_ke^{2iu\omega_0^1}=p_ke^{4iuJ\lambda_1}$.
In this way, we obtain
  \begin{align}
  \label{e20}
\frac{1}{N}\ln \frac{\chi(u)}{\chi_a(u)} \! = \! &\sum_{n=1}^{\infty} \!\frac{(-1)^{n+1}}{n}\frac{(e^{4iuJ\lambda_1}\!-\!1)^n}{2\pi}\! \int_{0}^{\infty}\mathrm \!\!\!\! dk
\: e^{-2n\pi J k^2/v}\nonumber \\
=&\frac{-v^{1/2}J^{-1/2}\sqrt{2}\mathrm{Li}_{3/2}(1-e^{4iuJ\lambda_1})}{8\pi},
  \end{align}
and
 \begin{gather}
  \begin{split}
\kappa_1=&N\left(\mu+\frac{v^{1/2}J^{1/2}\lambda_1\sqrt{2}}{2\pi}\right),\\
\kappa_2=&N\frac{v^{1/2}J^{3/2}\lambda_1^2(2\sqrt{2}-2)}{\pi},
  \end{split}
 \end{gather}
where $\mathrm{Li}_s(z)=\sum_{l=1}^{\infty}z^l/l^s$ is the polylogarithm
function. Also, we have
$f(u)=-\sqrt{2/J}\mathrm{Li}_{3/2}(1-e^{4Ju\lambda_1})/(8\pi)$ and
$\delta_n=1/2=d\nu/(z\nu+1)$. Obviously, the 1D transverse Ising model verifies our predictions in Eqs.~(\ref{e13},\ref{e15}).

We would also like to present some quantitative analysis about the work distribution function. $f(u)$ is a monotonic function with the following
asymptotic behavior: for $u\to -\infty$, $f(u)=\sqrt{2/J}\zeta(3/2)/(8\pi)$, where
$\zeta(z)$ is the Riemann zeta function; for $u\to \infty$ (the domain of $f(u)$
has been extended to the real axis by applying analytic continuation),
$f(u)=J(\lambda_1 u)^{3/2}\sqrt{2}/[\pi \Gamma(5/2)]$, where $\Gamma(z)$ is the
Gamma function. Hence, from the asymptotic behavior of $f(u)$ and by applying the Legendre-Fenchel transform, we have that
for $w<\mu$, $P_N(w)=0$ which is consistent with the initial ground state condition. A
confusion may arise when we consider $w>w_m\equiv\mu+\int_{0}^{\pi}\mathrm
dk\omega^1_{k}/\pi$ since now $P_N(w)$ is the probability of unphysical events.
This is a consequence of the approximation in which we extend the upper limit of
the integral to $\infty$. Actually, it can be proved that when $w>w_m$,
$I(w)>I(w_m)\propto v^{-1}$. Because $P_N(w)\propto e^{-NI(w)}$, $P_N(w)<P_N(w_m)\ll
1$, which indicates that the probabilities of the unphysical events are sufficiently small and our approximation is still reasonable.

If $\lambda_1$ is near the critical point, $\omega_k^1=2Jk$ when $k\to0$.
And for every mode, the dynamics corresponds to a half Landau-Zener
problem~\cite{tr1999,le2010,ad2006,dy2010}, where $p_k$ reads
 \begin{align}
  \label{e21}
  p_k=&1-2\frac{e^{-\pi\alpha_k/8}}{\pi\alpha_k}\sinh\left(\frac{\pi\alpha_k}{4}\right)\\
  &\times\left|\Gamma\left(1+\frac{i\alpha_k}{8}\right)+\sqrt{\frac{\alpha_k}{8}}\Gamma\left(\frac{1}{2}+\frac{i\alpha_k}{8}\right)e^{i\pi/4}\right|^2, \nonumber
 \end{align}
with $\alpha_k=4Jk^2/v$. This function has the following asymptotic behavior: for $\alpha_k\to0$, $p_k=1/2$; for $\alpha_k\to\infty$, $p_k=1/(2\alpha_k)^2$. Because only low-energy modes can get excited after the quench, we have
 \begin{align}
\kappa_1=&N\left(\mu+\frac{2J}{\pi}\int_{0}^{\infty}\mathrm dk \: k \: p_k\right)\approx N\left(\mu+0.038v\right)\\
\kappa_2=&\frac{8NJ^2}{\pi}\int_{0}^{\infty}\mathrm dk\: k^2\:p_k(1-p_k)\approx 0.092Nv^{3/2}J^{1/2}.\nonumber
\end{align}
For $n\geq 3$, the upper limit of the integral cannot be extended to $\infty$ due to the power-law decay of $p_k$. After some careful analysis, we find for $n>3$, $\kappa_n\sim v^2$ is reproduced~\cite{foot2}. Moreover, for $n=3$, the logarithmic correction appears: $\kappa_3\sim v^2\ln v$. These results again verify our predictions in Eq.~(\ref{e16}).

\textit{Conclusions}.---
In this Letter, we have studied the statistics of the work applied across
a quantum phase transition in systems characterized by independent excitations of quasiparticles.
We have shown that all the cumulants of the work distribution exhibit a scaling behavior for small quench rates, and that the scaling exponents are determined by
the dimension of the system and the critical exponents of the transition.
This is in analogy to the predictions of the KZM, although there are qualitative
differences in quenches ending close to and away from the critical point.
In addition, we are also able to determine the scaling exponents $\delta_n$ when (1) the energy spectrum
is always gapped during the protocol, (2) the initial state is not the ground state,
or (3) the protocol is a sudden quench protocol near the critical point (see supplemental material).
We also show that although the  cumulant CFW for slow linear quenches
traversing a critical point is analytic for $u \simeq 0$
(which allows to properly define the cumulants), it has non-analyticities
at certain values of $u$. This is related to the phenomenon of dynamical
quantum phase transitions (see supplemental material), which has been previously reported  for
the case of sudden quenches~\cite{dy2013, qu2016, st2008}.

H. T. Quan gratefully acknowledges support from
the National Science Foundation of China under grants
11775001, 11534002, and 11825001.
N. Freitas and M. Esposito acknowledge funding from the European Research
Council project NanoThermo (ERC-2015-CoG Agreement No. 681456).
V. Cavina is funded by the National Research Fund of Luxembourg
in the frame of project QUTHERM C18/MS/12704391.

\appendix

\section*{A: The regular analytic adiabatic perturbation theory}

If the energy spectrum is always gapped during the protocol, the scaling of the number of excitations
can be obtained by applying the regular adiabatic perturbation theory.
According to
Refs.~\cite{hi1988,be2008,le2010}, for a linear quench
$\lambda(t)=\lambda_0+v(t-t_0)$ with $v$ sufficiently small, the transition probability $P_{m,n}$
from the $m$th instantaneous eigenstate $|E^{\lambda_0}_m\rangle$ with
energy $E^{\lambda_0}_m$ to the $n$th instantaneous eigenstate
$|E^{\lambda_1}_n\rangle$ with energy $E^{\lambda_1}_n$ $(m\neq n)$ reads
 \begin{gather}
  \label{e31}
P_{m,n}=v^2\left[\frac{|\langle E^{\lambda_0}_n|\partial_{\lambda_0}|E^{\lambda_0}_m\rangle|^2}{(E^{\lambda_0}_n-E^{\lambda_0}_m)^2}+\frac{|\langle E^{\lambda_1}_n|\partial_{\lambda_1}|E^{\lambda_1}_m\rangle|^2}{(E^{\lambda_1}_n-E^{\lambda_1}_m)^2}
-2\cos(\gamma_{mn})\frac{\langle E^{\lambda_0}_n|\partial_{\lambda_0}|E^{\lambda_0}_m\rangle}{E^{\lambda_0}_n-E^{\lambda_0}_m}\frac{\langle E^{\lambda_1}_n|\partial_{\lambda_1}|E^{\lambda_1}_m\rangle}{E^{\lambda_1}_n-E^{\lambda_1}_m}\right],
 \end{gather}
where, without considering the Berry phase, we have $\gamma_{mn} = v^{-1}
\int_{\lambda_0}^{\lambda_1} d\lambda [E_m(\lambda) - E_n(\lambda)]$, and
in order to obtain a meaningful expansion, we require $v\langle
E^{\lambda}_n|\partial_{\lambda}|E^{\lambda}_m\rangle\ll
E^{\lambda}_n-E^{\lambda}_m$ for any $n\neq m$ and $\lambda$.
Eq.~(\ref{e31}) indicates that the probability of excitation is proportional to $v^2$ (we ignored the highly oscillating cosine function in Eq.~(\ref{e31})).
From this result, we can conclude, following the reasoning of Sec. III of the main text, that the scaling
exponents in the CFW are $\delta_n=2$. Moreover for later convenience, we also consider a sudden quench protocol with a sufficiently small amplitude $v$
\be
\lambda(t)=\lambda_0+v\: \Theta(t),
\ee
where $\Theta(t)$ is the Heaviside step function. As shown in Refs.~\cite{le2010,qu2010,qua2010}, we have
 \begin{gather}
  \begin{split}
  \label{es5}
P_{m,n}=&v^2\:|\langle E^{\lambda_0}_n|\partial_{\lambda_0}|E^{\lambda_0}_m\rangle|^2,
  \end{split}
 \end{gather}
and also in this case we obtain $\delta_n=2$.

We can demonstrate these results by considering the analytic solution for the cumulant CFW of a
forced harmonic oscillator, with the Hamiltonian given by
\be
\hat H(\lambda)=\frac{\hat{p}^2}{2m}+\frac{1}{2}m\omega^2\hat{x}^2+\lambda\sqrt{\frac{2m\omega}{\hbar}}\hat x.
\ee
According to Ref.~\cite{sta2008}, the CFW of this model satisfies the following equation
\be
\ln \frac{\chi(u)}{\chi_a(u)}=\left[(e^{iu\hbar\omega}-1)-\frac{4\sin^2(u\hbar\omega/2)}{e^{\beta\hbar\omega}-1}\right]A(\omega),
\ee
where $A(\omega)=\left|\int_{t_0}^{t_1}\mathrm ds \dot{\lambda}(s)e^{i\omega s}\right|^2$. It is easy to check that for the linear quench protocol,
\be
A(\omega)=\frac{2v^2}{w^2}\left\{1-\cos\left[\frac{(\lambda_1-\lambda_0)\omega}{v}\right]\right\},
\ee
and for the sudden quench protocol, $A(\omega)=v^2$, confirming in this way the quadratic scaling of the cumulant CFW with respect to $v$ for
both protocols.

\section*{B: The CFW for the independent excitations of  quasiparticles}

For independent excitations of quasiparticles, the work performed during the quench process reads
$w=N\mu+\sum_k\omega_kn_k$, where $n_k$
denotes the occupation number of the $k$th mode at the end of the process.
Let us call $p_{n_k}$ the discrete probability distribution of $n_{k}$ after the quench,
since $w$ is the sum of the independent random variables
$\omega_kn_k$, by additivity of the cumulant CFW, we can straightforwardly obtain
\be
\ln\chi(u)=Niu\mu+\sum_k \ln\chi_k(u),
\ee
where $\chi_k(u)=\sum_{n_k}e^{iu\omega_kn_k}p_{n_k}$ is the characteristic function of $\omega_kn_k$.
Notice that $p_{n_k}$ depends on the work
parameter $\lambda$, like the other relevant quantities $\mu$ and $\omega_k$.
Now, in a spirit similar to the few-excitation approximation, only the configurations in which at most one quasiparticle in each mode
gets excited at a time will be counted. Hence, for the excited mode $k$,
we let $p_{n_k=1}\approx p_k$, $p_{n_k=0}\approx1-p_k$  and $p_{n_k}\approx0$ for $n_k\geq 2$.
Thus, computing the CFW by adding up the contributions from all the configurations described above, we have

\be
\ln\chi(u)\approx Niu\mu+\sum_k \ln[1+p_k(e^{iu\omega_k}-1)].
\ee
Hence, according to Eq.~(9), we obtain the cumulants of work $\kappa_1\approx N\mu+\sum_k\omega_kp_k$ and $\kappa_n\approx\sum_k\omega_k^np_k$.
Here we have ignored the quantity $p_k^n$ for $n\geq 2$, which does not influence
the scaling behavior of the CFW.

\section*{C: The exponents in the CFW for an initial state other than the ground state}

The scaling relations given by Eq.~(1) are the consequence of the vanishing energy gap (due to the excitation energies of the low-energy modes approach zero) at the quantum critical point.
With this in mind, the critical behavior is independent of the choice of the initial state and is expected to hold
also when initializing our quantum system in an excited configuration.
Since the validity of the KZM and the adiabatic perturbation theory is preserved in this case,
we can apply the same reasoning of sections II and III of the main text to find that
$\delta_n$ are exactly the same as the ones for the initial ground state.
In addition, the above results are also valid for the
canonical initial state, given that $\lambda_0$ is sufficiently far from the critical point.
When the temperature is high and $\lambda_0$ is near the critical point, on the contrary,
$v^{\delta_n}$ can achieve larger (for bosons) or smaller (for fermions) values than the ones derived in the main text~\cite{le2010}.

As a technical remark, it is worth mentioning that the adiabatic part of the cumulant CFW, $\ln\chi_a(u)$,
in the cases considered above should be defined as the contribution to the cumulant CFW due to the
adiabatic evolution of the corresponding initial state (that in the present discussion is not necessarily
the ground state).

We can demonstrate our results by considering a
1D transverse Ising model initialized in a canonical distribution.
If both $\lambda_0$ and $\lambda_1$ are away from the critical point, due to the exponential
decay of $p_k$, only low-energy modes can get excited after the quench
and we have $p_kh(\omega_{k}(\lambda))\approx p_kh(\omega_{0}(\lambda))$,
where $h(\omega_{k}(\lambda))$ is an arbitrary smooth function of $\omega_k(\lambda)$. Then, according to Eqs.~(19, 20, 21), we obtain
 \begin{gather}
  \begin{split}
  \label{e27}
\ln\chi_a(u)&=\frac{N}{2\pi}\int_{0}^{\pi} \mathrm dk \ln\frac{1+\cos(u\omega_k^1-u\omega_k^0+i\beta\omega_k^0)}{1+\cosh(\beta\omega_k^0)}\\
\frac{1}{N}\ln\frac{\chi(u)}{\chi_a(u)}&=
\frac{-v^{1/2}J^{-1/2}\sqrt{2}}{8\pi}\mathrm{Li}_{3/2}\left(\frac{-2\sin(2J\lambda_1u)\sin[2J\lambda_0(u-i\beta)]}{1+\cos[2J(\lambda_1u+\lambda_0u-i\lambda_0\beta)]}\right)\\
\kappa_1 &=N\mu+ N \frac{v^{1/2}J^{1/2}\lambda_1\sqrt{2}}{2\pi}\tanh(\beta J\lambda_0)\\
\kappa_2 &=\kappa_{2a}+N\frac{v^{1/2}J^{3/2}\lambda_1[(2\sqrt{2}-2)\lambda_1\sinh^2(\beta J\lambda_0)-\sqrt{2}\lambda_0]}{\pi\cosh^2(\beta J\lambda_0)},
  \end{split}
 \end{gather}
where the first and the second cumulants of work under the adiabatic driving are given by
 \begin{gather}
  \begin{split}
  \mu =&\frac{1}{2\pi}\int_{0}^{\pi} \mathrm dk (\omega_k^0-\omega_k^1)\tanh\left(\frac{\beta\omega_k^0}{2}\right)\\
  \kappa_{2a}=&\frac{N}{4\pi}\int_{0}^{\pi} \mathrm dk (\omega_k^0-\omega_k^1)^2\mathrm{sech}^2\left(\frac{\beta\omega_k^0}{2}\right).
  \end{split}
\end{gather}
As we have mentioned above, the exponents are $\delta_n = 1/2$, just the same as that for
the initial ground state (Eqs.~(24,25)).

If $\lambda_0$ is near the critical point and $\lambda_1$ is away from it,
$\omega_k^0=2Jk$ when $k\to0$ and $p_k$ is given by Eq.~(26) because we fall again in the
half Landau-Zener scenario~\cite{le2010,ad2006,dy2010}.
Thus, in the high temperature limit, since only low-energy modes can get
excited after the quench, we have
 \begin{gather}
  \begin{split}
  \label{ec3}
\frac{1}{N}\ln \frac{\chi(u)}{\chi_a(u)}&=\frac{2J}{\pi}(i\beta-u)\tan(J\lambda_1u)\int_{0}^{\infty}\mathrm dk \:k \:p_k
=0.038v(i\beta-u)\tan(J\lambda_1u)\\
\kappa_1&=N\left(\mu+0.038\beta vJ\lambda_1\right)\\
\kappa_2&=\kappa_{2a}+0.076NvJ\lambda_1.
  \end{split}
 \end{gather}
Comparing Eq.~(\ref{ec3}) with Eq.~(\ref{e27}), we recognize that the scaling $v^{1/2}$ for the initial ground state
becomes $\beta v$ for the odd-order cumulants of work
and to $v$ for the even-order cumulants of work. This is compatible with the fermionic
anti-bunching effects discussed in Ref.~\cite{le2010}, since the quasiparticles in
the transverse Ising model are fermions.

\section*{D: Sudden quench protocol near the critical point}

In this section, we discuss the scaling behavior of the cumulant CFW for the sudden quench protocol near the critical point
\be
\label{ed1}
\lambda(t)=v\:\Theta(t).
\ee
If the quench amplitude $v$ is sufficiently small, the adiabatic perturbation theory is still valid~\cite{le2010,qu2010,qua2010}
and we can write
\be
p_k\approx\left|\int_{0}^{v}\frac{\mathrm d\lambda}{\lambda}\: G\left(\frac{k}{\lambda^\nu}\right) \right|^2,
\ee
where $G(x)$ is the scaling function defined in Eq.~(6) of the main text.
To get rid of the dependence on $v$ in the upper limit of the integral,
we can define the following rescaled quantities: $\lambda=\theta v$, $k=\phi v^{\nu}$
and finally obtain, from Eq.~(9) of the main text,
\begin{equation}
\label{es14}
v^{\delta_n}=\left\{
\begin{aligned}
&v^{(d+nz)\nu}\ \ \ &(d+nz)\nu<2\\
&v^2\ln v  &(d+nz)\nu=2\\
&v^2   &(d+nz)\nu>2,
\end{aligned}
\right.
\end{equation}
in which the exponents explicitly depend on $n$. Please note that results in Eq.~(\ref{es14}) are different from those in Eq.~(\ref{es5}).

Again, we use the 1D transverse Ising model to demonstrate our results. According to Ref.~\cite{gr2019}, $p_k$ is such that
 \begin{gather}
1-2p_k=
\frac{(v-1+\cos k)(-1+\cos k)+\sin^2 k}{\sqrt{[(v-1+\cos k)^2+\sin^2 k][(-1+\cos k)^2+\sin^2 k]}}.
 \end{gather}
Although $v$ is small, we cannot directly expand the equation above in Taylor series since it will result
in a divergence of the excitation probability for $k=0$ and make the integral in Eq.~(23) ill-defined.
This also happens to the fidelity susceptibility  at the critical point of a transverse Ising model~\cite{fi2013,sc2009}.
To obtain the proper integral, we have to first rescale $k=k'v$ and then perform the series expansion.
With this in mind, for low energy modes, we have
\be
p_k\approx\frac{\sqrt{1+k'^2}-k'}{2\sqrt{1+k'^2}}=\frac{\sqrt{v^2+k^2}-k}{2\sqrt{v^2+k^2}}\in[0,1].
\ee
Since $\omega_k^1=2Jk$ for low energy modes, we obtain the first and the second cumulant of work~\cite{foot4}
 \begin{gather}
  \begin{split}
  \label{ec6}
\kappa_1=&N\left(\mu+\frac{2J}{\pi}\int_{0}^{\pi}\mathrm dk k p_k\right)=N\left(\mu-\frac{J}{2\pi}v^2\ln v\right)\\
\kappa_2=&\frac{8NJ^2}{\pi}\int_{0}^{\pi}\mathrm dkk^2p_k(1-p_k)=2Nv^{2}J^{2}.
  \end{split}
 \end{gather}
Also, for $n\geq 3$, $\kappa_n\sim v^2$ is recovered. Thus, these results verify Eq.~(\ref{es14}).

\section*{E: Dynamical quantum phase transition}

Dynamical quantum phase transitions are associated with the nonanalytic behavior of
physical quantities as functions of time during the evolution of a  quantum
system~\cite{dy2013,qu2016}.
In this dynamical process, the time acts as a controlling parameter and the time domain
is partitioned into different regions, in which the physical quantities are characterized by qualitatively different
behaviors.
Such a phenomenon can be characterized, for instance, by the rate function of the Loschmidt
echo of the 1D transverse Ising chain. It is nonanalytic
at critical times~\cite{dy2013,qu2016} when considering the thermodynamic limit.
The dynamical quantum phase transitions are also relevant in the characterization of the work fluctuations.
In fact, as Refs.~\cite{dy2013,qu2016,st2008}
clarified, a connection between the Loschmidt echo and the work distribution (or its characteristic function) can be
established at least in
two cases: for a single~\cite{qu2016,st2008} and for a double~\cite{dy2013,qu2016} sudden quench protocol.

Although the quench considered in our letter is linear, we also find
nonanalytic behaviors of the CFW $\chi(u)$ for large $u$. Since the variable $u$ can be
interpreted as the time of the Loschmidt echo in Refs.~\cite{qu2016,st2008}, we
still refer to this phenomenon as a dynamical quantum phase transition.
For the 1D
transverse Ising model, the nonanalytic behavior occurs when the integrand in
Eq.~(22) is divergent,  i.e., when $1+p_k(e^{2iu\omega_k^1}-1)=0$ .
Then, we obtain the critical times $u^*$
\be
u^*=\frac{\pi}{2\omega^1_{k^*}}\left(n+\frac{1}{2}\right),\ \ \ n=0,\pm1,\cdots,
\ee
and the critical mode $k^*=\sqrt{v\ln 2/(2\pi J)}$, since $p_{k^*}=1/2$.
However, the integral is still well-defined at the critical times by
applying analytic continuation. Actually, the cumulant CFW $\ln\chi(u)$
and its derivative are discontinuous at $u^*$, i.e.,
$\lim_{u\to
u^{*-}}\ln[\chi(u)/\chi_a(u)]=-v^{1/2}\sqrt{2/J}\mathrm{Li}_{3/2}(2)/(8\pi)$,
$\lim_{u\to
u^{*+}}\ln[\chi(u)/\chi_a(u)]=-v^{1/2}\sqrt{2/J} \: \: \overline{\mathrm{Li}_{3/2}(2)}/(8\pi)$,
$\lim_{u\to
u^{*-}}\partial_u\ln[\chi(u)/\chi_a(u)]=-v^{1/2} \sqrt{2J} \lambda_1 \mathrm{Li}_{1/2}(2)/(4\pi)$
and $\lim_{u\to
u^{*+}}\partial_u\ln[\chi(u)/\chi_a(u)]=-v^{1/2}\sqrt{2J}\lambda_1\overline{\mathrm{Li}_{1/2}(2)}/(4\pi)$,
where $\overline{z}$ denotes the complex conjugate of $z$. It is worth
mentioning that these critical times exist only when the protocol goes through
the critical point because without crossing the critical point (e. g., $\lambda_0<\lambda_1<0$), we always have $p_k<1/2$ for all the modes.


\begin{thebibliography}{99}
\bibitem{foot3} For $n\leq 3$, $\kappa_n$ is equal to the $n$th central moment of work.
\bibitem{foot1} The condition for the convergence of the series $2p_k|\sin(u\omega_k^1)|<1$ is always satisfied in Eq.~(\ref{e19}) when we calculate the cumulants of work, because $u\to 0$ in this case.
\bibitem{foot2} In fact, for $n>3$, Eq.~(\ref{e21}) overestimates the results compared with numerical results because for high-energy modes, the calculation of $p_k$ does not exactly correspond to a half Landau-Zener problem. However, Eq.~(\ref{e21}) does lead to the correct scaling behavior of the cumulant CFW.
\end{thebibliography}
\end{document}